\documentclass[onecolumn,showkeys,showpacs,preprintnumbers,amsmath,amssymb,superscriptaddress,aps,prc]{revtex4}
\usepackage{graphicx}
\usepackage{xcolor}
\usepackage{subfigure}

\begin{document}

\title{Dipole-dipole dispersion interactions between neutrons}
\author{James F. Babb}
\affiliation{ITAMP, Harvard-Smithsonian Center for Astrophysics, MS 14, 60 Garden St., Cambridge, MA 02138, USA}
\author{Renato Higa}
\affiliation{Instituto de F\'{\i}sica,
Universidade de S\~{a}o Paulo, R. do Mat\~{a}o 1371, 05508-090, S\~{a}o Paulo,
Brazi}
\author{Mahir S. Hussein}
\affiliation{Instituto de Estudos Avan\c{c}ados, Universidade de S\~{a}o Paulo C. P.
72012, 05508-970 S\~{a}o Paulo-SP, Brazil,\\
 Instituto de F\'{\i}sica,
Universidade de S\~{a}o Paulo, R. do Mat\~{a}o 1371, 05508-090, S\~{a}o Paulo,
Brazil,\\
  Departamento de F\'{i}sica, Instituto Tecnol\'{o}gico de Aeron\'{a}utica, CTA, S\~{a}o Jos\'{e} dos Campos, S.P., Brazil}

\begin{abstract}
We investigate the long-range interactions between two neutrons 
utilizing recent data on the neutron static and dynamic electric and magnetic dipole polarizabilities.
The resulting long-range potentials are used to make  quantitative comparisons
between the collisions of a neutron with a neutron and a neutron with a proton. 
We also assess the importance of the first pion production threshold and first 
excited state of the nucleon, the $\Delta$-resonance 
($J^{\pi}$ = + 3/2, I = 3/2). 
We found both dynamical effects to be quite relevant for distances $r$ between 
$\sim 50$~fm up to $\sim 10^3$~fm in the nn system, the neutron-wall system and 
in the wall-neutron-wall system, reaching the expected asymptotic limit 
beyond that. 
Relevance of our findings to the confinement of ultra cold neutrons inside bottles is discussed.
\end{abstract}

\pacs{14.20.Dh, 25.40.Dn, 25.40.Cm}
\keywords{Scattering theory, Casimir-Polder interaction. Neutron-neutron system, neutron-wall interaction}

\maketitle

\section{Introduction}
A polarizable particle is expected to exhibit a long-range electromagnetic interaction.
Examples include 
the charge-induced dipole interaction energy between an electron and a hydrogen atom, $-\frac{1}{2}\alpha e^2/d^4$,
and the dipole-dipole dispersion  interaction energy between two hydrogen atoms,
$-C_6 / d^6$, where $d$ is the separation  
between the electron and the H atom or between the two H atoms, $\alpha = \frac{9}{2} a_0^3$ 
is the atomic electric dipole polarizability,
and $C_6 \approx 6.5 e^2a_0^5$ 
is the dispersion or van der Waals constant.
The quantities $\alpha$ and $C_6$ can be expressed in terms of the electric dipole oscillator
strength distribution of the atom, which describes the response of the electron to photons at specific frequencies
of an external electric field.
Higher order interaction energies involving magnetic dipole and higher-order multipolar
polarizabilities, and multipolar dispersion constants are also well-characterized and these
may be expressed in terms of the corresponding multipolar oscillator strengths.
These interaction potentials are general for atoms and molecules  
and they are widely studied and applied for descriptions of spectroscopy and scattering.

The neutron possesses internal structure (two quarks down and one quark up) 
with an electric dipole polarizability $\alpha_n$ and a magnetic dipole 
polarizability $\beta_n$, usually viewed as the response of the pion cloud 
to external electromagnetic 
fields~\cite{SchRauRie89,WisLevSch98,GriMcGPhi12,HolSch14}.
The most recent recommended values from the Particle Data Group (PDG) are
$\alpha_n = (11.6 \pm1.5) \times 10^{-4}$ fm${}^{3}$  and $\beta_n = (3.7 \pm2.0) \times 10^{-4}$ fm${}^{3}$~\cite{pdg}.
Another recommendation is $\alpha_{n}= (12.5 \pm 1.8) \times 10^{-4}\; \textrm{fm}^3$ 
and $\beta_{n} = (2.7 \mp 1.8) \times 10^{-4}\; \textrm{fm}^3$~\cite{KosCamWis03}. (The appearance of $\pm$ and $\mp$ is related to the sum rule 
used to determine these values.) 
A number of separate groups determined the neutron electric dipole polarizability $\alpha_n$ 
by measuring the effect of the potential energy
\begin{equation}
 -\textstyle{\frac{1}{2}} \alpha_n e^2Z^2 / R^4,
\end{equation}
where $Z$ is the nuclear charge and $R$ is the separation distance,
on the scattering amplitude~\cite{Tha59} in neutron scattering by ${}^{208}\textrm{Pb}$ nuclei.
The high nuclear charge generates an electric field that polarizes the neutron leading
to an effect completely analogous and of the same form as that mentioned in
the first paragraph for the charge-induced 
dipole interaction between an electron and an atom.
Experiments were carried out looking at the differential scattering of neutrons on Pb and
by looking at neutron transmission through Pb.
For example, using neutron transmission through lead Schmiedmayer \textit{et al.}~\cite{SchRieHar91} obtained 
the value $\alpha_n = (12.0 \pm1.5 \pm 2.0) \times 10^{-4}$ fm${}^{3}$,
where the first uncertainty is statistical and the second uncertainty is systematical.
Further analyses of this experiment and
discussions of other neutron-nucleus
scattering experiments can be found in Refs.~\cite{Ale96,WisLevSch98,Pok00}.
Two other experimental approaches to the neutron polarizability are through measurements of
quasi-free Compton scattering from the bound 
neutron in the deuteron ($\gamma + \textrm{d(np)} \rightarrow \gamma' + \textrm{d(np)}$) and of elastic photon scattering
from the deuteron ($\gamma+\textrm{d}\rightarrow \gamma +\textrm{d}$)~\cite{DrePasVan03,Hagelstein:2015egb}, 
where the observable quantities are the Compton polarizabilities
$\bar{\alpha}_n + \bar{\beta}_n$
and $\bar{\alpha}_n- \bar{\beta_n}$. In particular, Compton scattering  experimental results, with the use
of sum rules,  led to the recommended values for $\beta_n$, cited above.
One can also use sum rules and the result for $\alpha_n$ from
neutron scattering experiments to determine 
$\beta_n$~\cite{SchRieHar91,Hagelstein:2015egb}.
Compton scattering implies a response of the neutron to photon energies and
provides a connection to the polarizabilities.
Care is required in using theoretical Compton scattering frequency-dependent amplitudes because
of the presence of relatively small corrections arising from the relativistic
wave equations utilized~\cite{Pet64,KarMil99}, but
no conceptual difficulties arise in relating
static polarizabilities arising in Compton scattering
to those arising from an external electric field because they are identical~\cite{KosCamWis03,Sch05}.
In the past few years the frequency-dependent values of $\alpha_n$ and 
$\beta_n$ have been calculated in the framework of chiral effective field 
theory~\cite{Hil05,GriMcGPhi12,Lensky:2012ag,Hagelstein:2015egb}, the 
effective theory of the underlying quantum chromodynamics (QCD). Guided by 
the approximate chiral symmetry of QCD, these calculations show good 
agreement with data, though convergence patterns are different depending 
on its covariant or heavy-baryon formulations, and on the explicit inclusion 
of the Delta resonance (c.f. Ref.~\cite{Lensky:2012ag}).

The paper addresses the influence of the internal structure of the neutron and proton on their dispersive interactions 
with another neutron. In the second section we look at the neutron-neutron Casimir-Polder (CP) interaction and in the third 
section we compare the CP effect in the nn and pn systems. In the fourth section we derive the CP interaction in the 
neutron-wall system, and finally in the fifth section, we give our concluding remarks. 

\section{The neutron-neutron Casimir-Polder interaction}
We believe it is reasonable to expect that there is a neutron-neutron dispersion interaction, 
based on the experimental evidence for the static polarizabilities $\alpha_n$ and $\beta_n$ and
on theoretical calculations of the polarizabilities as functions of photon energies.
Indeed, while our conclusion is based on a physical
analogy between the neutron and the H atom
there is a more formal basis for such an expectation.
Feinberg and Sucher showed that 
for asymptotically large separations, retarded dispersion interactions
between two ``systems'' (electromagnetically polarizable particles) 
result independently of the system models and follow
from general principles of Lorentz invariance, electromagnetic current conservation,
analyticity, and unitarity~\cite{FeiSuc70,FeiSuc79}.
An early application of these ideas to a calculation of the neutron-neutron
scattering length was carried out by Arnold~\cite{Arn73}.
However, at the time of 
his analysis $\beta_n$ was unknown, the accepted value of $\alpha_n$ was twice today's value, 
and dynamic polarizabilities were unavailable.

Following Feinberg and Sucher~\cite{FeiSuc70}, the asymptotic 
($r\sim\infty$) long-range interaction potential between two neutrons is
given by the Casimir-Polder potential
\begin{equation}
\label{nn-infty}
V^\infty_{CP,nn}(r) =  - (\hbar c/4\pi)[23(\alpha_{n}^2 + \beta_{n}^2)
 -14\alpha_{n}\beta_{n} ]{r^{-7}} + \mathcal{O}(r^{-9})
=V^{*}_{CP,nn}(r) + \mathcal{O}(r^{-9}),
\end{equation}
with the notation $V^{*}_{CP}$ meaning the static limit of the nucleon 
dynamic polarizabilities. 
In contrast, the asymptotic long-range interaction potential between a proton
and a neutron is the sum of the charge-induced dipole interaction potential
and the Casimir-Polder-type potential for a neutral polarizable particle and
a charged particle~\cite{BerTar76} 
\begin{equation}
\label{np-infty}
V^\infty_{CP,pn}(r) = \hbar c\,\alpha_0\left[ 
-\frac{1}{2}\alpha_{n}r^{-4}
+\frac{1}{4 \pi c M_p}(11\alpha_n + 5\beta_n) r^{-5} 
+ \mathcal{O}(r^{-7}) \right]
=V^{*}_{CP,pn}(r)  + \mathcal{O}(r^{-7}),
\end{equation}
where $M_p$ is the proton mass and $\alpha_0=e^2/4\pi\sim 1/137$ is the 
electromagnetic fine structure constant.
Note the appearance of a repulsive $r^{-5}$ potential for
the asymptotic interaction of a neutron and a proton.
(We would expect the polarizability of the proton to enter at 
the higher order of $ \mathcal{O}(r^{-7})$ through
a potential similar to Eq.~(\ref{nn-infty}) that is bilinear in neutron and proton polarizabilities~\cite{FeiSuc70}.)

Accordingly, estimates that improve on 
the asymptotic Casimir-Polder interaction between two neutrons 
and between a neutron and a proton can be obtained from, respectively, Eqs.~(\ref{nn-infty}) and (\ref{np-infty}),
where we have substituted the accepted polarizability values and converted the expressions
to suitable units for nuclear physics,
\begin{equation}
\label{nn-infty-num}
V^{*}_{CP,nn}(r) \approx  - 0.49 \times 10^{-3} (r/\textrm{fm})^{-7}  \;\textrm{MeV}, \quad r\sim\infty ,
\end{equation}
and 
\begin{equation}
\label{np-infty-num}
V^{*}_{CP,pn}(r) \approx  0.91 \times 10^{-3} (r/\textrm{fm})^{-4} [-1 + 0.40 (r/\textrm{fm})^{-1}]  \;\textrm{MeV}, \quad r\sim\infty .
\end{equation}
More generally, The Casimir-Polder theory gives the interaction between two identical neutral
polarizable particles  valid for all distances sufficiently large that 
exchange forces are negligible~\cite{FeiSuc70,Bab10},
\begin{equation}
\label{CP-dip-dip}
V_{CP,ij}(r) = - \frac{\alpha_0}{\pi r^6} I_{ij}(r)
\end{equation}
where
\begin{eqnarray}
\label{eq:integ_ij}
&&I_{ij}(r) = \int_{0}^{\infty} d\omega e^{- 2\alpha_{0} \omega r} \Big\{
\big[\alpha_i(i\omega)\alpha_j(i\omega)+\beta_i(i\omega)\beta_j(i\omega)
\big]P_E(\alpha_{0} \omega r)
\nonumber\\[1mm]&&
\hspace{2.5cm}
+\big[\alpha_i(i\omega)\beta_j(i\omega)+\beta_i(i\omega)\alpha_j(i\omega)
\big]P_M(\alpha_{0} \omega r)
\Big\},
\nonumber\\[3mm]&&
P_E(x) = x^4 + 2x^3 + 5x^2 + 6x + 3 ,
\qquad
P_M(x) = - ( x^4 + 2x^3 + x^2 ) ,
\end{eqnarray}
$\alpha_i(\omega)$ and $\beta_i(\omega)$ are respectively the dynamic electric 
and magnetic dipole polarizability of particle $i$, similarly for 
particle $j$ 
\footnote{We note that an analogue of Eq.~(\ref{eq:integ_ij}) was recently
derived for \textit{gravitational} interactions: The leading term involves
dynamic gravitational quadrupole polarizabilities~\cite{FHK16}. 
The connection between electromagnetic and gravitational polarizabilities 
is also discussed in~\cite{Holstein17}.}. 
Detailed analyses of the neutron based on chiral effective field theory, 
for photon energies up to the excitation of the $\Delta$ resonance 
are found in Refs.~\cite{Hildebrandt:2003fm,Hil05,GriMcGPhi12,Lensky:2015awa,
Hagelstein:2015egb}. 
The analytic expressions for the neutron polarizabilities are far from simple. 
However, we attempt to 
parametrize $\alpha_n(\omega)$ and $\beta_n(\omega)$ in terms of relatively 
simple formulas that incorporate the most important low-energy aspects. 

\begin{figure}[tbh]
\begin{center}
\includegraphics[width=0.45\textwidth,clip]{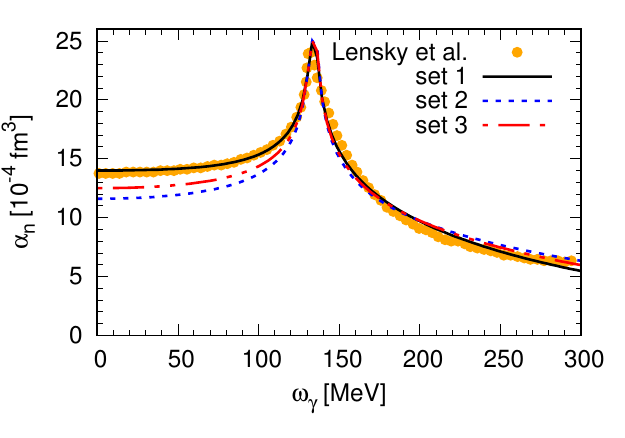}
\includegraphics[width=0.45\textwidth,clip]{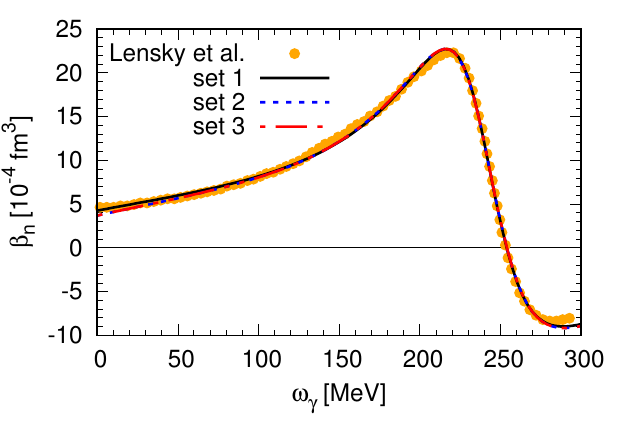}
\caption{\protect 
Dynamic electric (left) and magnetic (right) polarizabilities, as functions 
of the photon energy $\omega_{\gamma}$.}
The yellow circles are the CB-$\chi$EFT results of Lensky 
{\em et al.}~\cite{Lensky:2015awa} while sets 1, 2, and 3 correspond to 
our parametrizations using the numbers specified in Table~\ref{tab:pol-par}.
\label{fig:n-polariz}
\end{center}
\end{figure}

Our parametrization for the dynamic electric dipole polarizability reads 
\begin{equation}
\label{eq:edip-polariz1}
\alpha_{n} (\omega) = 
\frac{\alpha_n(0)\,\sqrt{(M_{\pi}+a_1)(2M_{n}+a_2)}(0.2a_2)^2}
{\sqrt{(\sqrt{|M_{\pi}^2-\omega^2|}+a_1)(\sqrt{|4M_{n}^2-\omega^2|}+a_2)}
\big[|\omega|^2+(0.2a_2)^2\big]}\,.
\end{equation}
Besides the static electric polarizability $\alpha_n(0)$ and the masses of 
the pion ($M_{\pi}$) and the neutron ($M_{n}$), this expression has two 
mass parameters $a_1$ and $a_2$. 
The parameter $a_1$ is formally a higher-order effect, but important 
to match the correct pion production threshold, which controls the low-energy 
behavior of $\alpha_n(\omega)$~\cite{Hildebrandt:2003fm}. 
The square roots in Eq.~(\ref{eq:edip-polariz1}) are an attempt to 
incorporate the pion production threshold behavior, above which 
$\alpha_n$ develops an imaginary part. 
This specific form also assumes a smooth and asymptotically decreasing 
behavior of $\alpha_n$ at 
imaginary frequencies, which is expected from analyticity of the 
Compton $S$-matrix and used in the construction of our Casimir-Polder 
potentials. 
We fit the above expression to the curves of Lensky, McGovern, and 
Pascalutsa~\cite{Lensky:2015awa}, 
results from the covariant formulation of baryon chiral effective field 
theory (CB-$\chi$EFT). In contrast to the non-relativistic, heavy baryon 
formulation of $\chi$EFT (HB-$\chi$EFT), the former properly takes into 
account recoil corrections to all orders, which is relevant to correctly 
describe the threshold behavior due to pion production. 
For $M_n=938.919$~MeV we obtain 
$M_{\pi}=134.051$~MeV, fairly close to the neutral pion mass (134.98 MeV). 
The remaining 
parameters are presented in Table~\ref{tab:pol-par}. In Set 1 we let 
$\alpha_n(0)$ be a fit parameter, in Set 2 we keep $\alpha_n(0)$ fixed to 
the PDG central value~\cite{pdg}, and in Set 3 we keep $\alpha_n(0)$ fixed 
to the central value of Ref.~\cite{KosCamWis03}. 
The quality of the parametrization can be 
observed on the left panel of Fig.~\ref{fig:n-polariz}, well within the 
expected theoretical uncertainties (see 
Refs.~\cite{Lensky:2015awa,Hagelstein:2015egb}).

For the dynamic magnetic dipole polarizability we use 
\begin{equation}
\label{eq:mdip-polariz1}
\beta_{n} (\omega) = 
\frac{\beta_n(0)-b_1^2\omega^2+b_2^3{\rm Re}(\omega)}
{(\omega^2-\omega_{\Delta}^2)^2+|\omega^2\Gamma_{\Delta}^2|}\,,
\end{equation}
which incorporates the relevant physics in this quantity, namely, the 
$\Delta$ resonance. In fact, from the fit parameters $b_1$, $b_2$, 
$\omega_{\Delta}$, and $\Gamma_{\Delta}$, the last two are close to 
the $n$-$\Delta$ mass splitting~\footnote{More precisely, this value is closer 
to $M_n-M_{\Delta}-\Gamma_{\Delta}/2$, the onset of the $\Delta$ resonance 
contributions.} 
and $\Delta$ resonance width, respectively. 
The term proportional to ${\rm Re}(\omega)$ mimics the onset of an 
imaginary term in the Compton amplitude above the real photon threshold, 
that would otherwise be absent below it. 
As in the $\alpha_n$ case, this specific form assumes a smooth and 
asymptotically decreasing behavior of $\beta_n$ at imaginary frequencies. 
The fitted parameters are shown in Table~\ref{tab:pol-par}, with $\beta_n(0)$ 
evaluated in analogous way as $\alpha_n(0)$ for each Set. 
One observes very little spread for this quantity among these three different 
sets. However, we noticed numerically that the contribution of 
$\beta_n(\omega)$ amounts to a decrease of about 10\% in the strength of 
the Casimir-Polder interaction between two neutrons, $V_{CP,nn}(r)$. This 
CP potential is, therefore, most sensitive to the differences observed in the 
description of $\alpha_n(\omega)$, as we discuss later. 
\begin{table}[tbh]
\begin{center}\begin{tabular}{|c||c|c|c|c|c|c|c|c|}\hline
& $\alpha_n(0)$ ($10^{-4}{\rm fm}^3$) & $a_1$ (MeV) & $a_2$ (MeV) & 
$\beta_n(0)$ ($10^{-4}{\rm fm}^3$) & $b_1$ (MeV) & $b_2$ (MeV) & 
$\omega_{\Delta}$ (MeV) & $\Gamma_{\Delta}$ (MeV) \\ \hline
Set 1 & 13.9968 & 12.2648 & 1621.63 & 4.2612 & 8.33572 & 22.85 & 241.484 & 66.9265 \\ \hline
Set 2 & 11.6 & 2.2707 & 2721.47 & 3.7 & 8.67962 & 24.2003 & 241.593 & 68.3009 \\ \hline
Set 3 & 12.5 & 5.91153 & 2118.79 & 2.7 & 9.27719 & 26.328 & 241.821 & 70.8674 \\ \hline
\end{tabular}\end{center}
\caption{Parameters of Eqs.~(\ref{eq:edip-polariz1}), (\ref{eq:mdip-polariz1}) 
fitted to the theoretical curves of Ref.~\cite{Lensky:2015awa}. 
See text for details. 
\label{tab:pol-par}}
\end{table}

In order to assess the quality of Eqs.~(\ref{eq:edip-polariz1}), 
(\ref{eq:mdip-polariz1}) at imaginary frequencies, we compared them to the 
heavy-baryon chiral EFT (HB-$\chi$EFT) expressions given by Hildebrandt 
{\em et al.}, Appendices B and C of Ref.~\cite{Hildebrandt:2003fm}. 
We made sure to reproduce their results at real $\omega$, then extrapolated 
to the imaginary domain. HB-$\chi$EFT lies between our set 2 and set 3 with 
Eq.~(\ref{eq:edip-polariz1}) up to about $i\omega=iM_{\pi}$. 
On the other hand, for the magnetic case Eq.~(\ref{eq:mdip-polariz1}) starts 
disagreeing with HB-$\chi$PT beyond 
$i\omega\approx i\,\mathrm{50\,MeV}\sim iM_{\pi}/3$. However, we checked 
numerically that the magnetic contribution to the Casimir-Polder potentials 
is at most a 15\% effect. 
We also noticed that the HB-$\chi$EFT results for $\alpha_n(i\omega)$ and 
$\beta_n(i\omega)$ exhibit a numerical 
singularity as one approaches $i\omega=iM_{\pi}$. In such complex-$\omega$ 
domain the non-Born Compton amplitudes, from which $\alpha_n(\omega)$ and 
$\beta_n(\omega)$ are obtained, should not exhibit any low-energy physical 
singularities. This is probably a consequence of the heavy-baryon formalism 
in missing the correct pion-production threshold, which is normally fixed 
``by hand''~\cite{Hildebrandt:2003fm,Lensky:2015awa,Hagelstein:2015egb}. 
In this exploratory work we rely on our parametrizations 
(\ref{eq:edip-polariz1}), (\ref{eq:mdip-polariz1}), with room for technical 
improvements postponed to future works. 

\begin{figure}[tbh]
\begin{center}
\includegraphics[width=0.65\textwidth,clip]{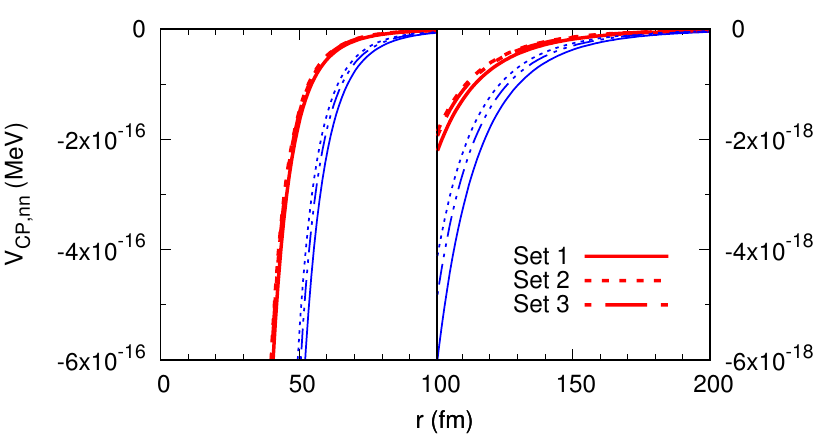}
\caption{\protect 
CP-interaction for two neutrons, as a function of the separation distance $r$. 
The red thick and blue thin lines correspond to the use of dynamical and 
static dipole polarizabilities, respectively.}
\label{fig:Vcp_nn01}
\end{center}
\end{figure}

Given the dynamic electric~(\ref{eq:edip-polariz1}) and 
magnetic~(\ref{eq:mdip-polariz1}) polarizabilities, one computes the 
neutron-neutron CP-interaction via Eqs.~(\ref{CP-dip-dip}) and 
(\ref{eq:integ_ij}). 
Fig.~\ref{fig:Vcp_nn01} shows the CP-interaction for two neutrons, 
$V_{CP,nn}(r)$, as a function of the separation distance. 
The bold red curves correspond to $V_{CP,nn}(r)$ given by the dynamic 
polarizabilities previously shown, while the thin blue curves correspond 
to the static limit $\alpha_n(\omega)$, $\beta_n(\omega)$ $\to$ 
$\alpha_n(0)$, $\beta_n(0)$. In such limit, integration of 
Eq.~(\ref{eq:integ_ij}) is straightforward and leads to Eq.~(\ref{nn-infty}). 

\begin{figure}[tbh]
\begin{center}
\includegraphics[width=0.55\textwidth,clip]{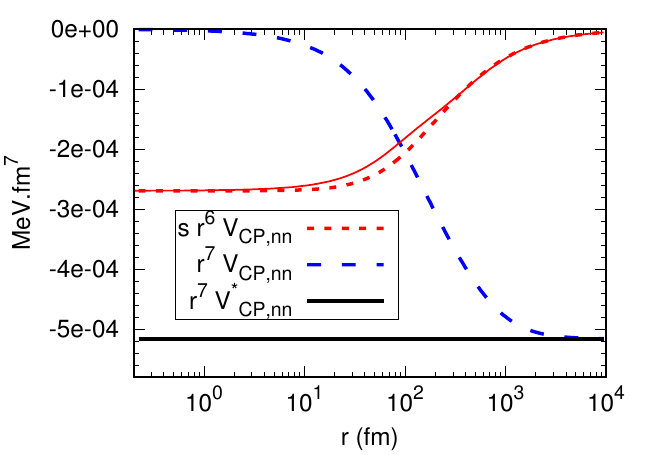}
\caption{\protect 
The neutron-neutron CP-interaction as a function of the separation distance 
$r$, multiplied by $s\,r^6$ (red dotted line, with $s=100$~fm) and $r^7$ 
(blue long-dashed line). The black solid line is the CP-potential from 
the static limit of the dipole polarizabilities, multiplied by $r^7$.}
\label{fig:Vcp_nn02}
\end{center}
\end{figure}

The difference between the use of dynamic and static polarizabilities is 
evident from the curves. In order to assess the expected long-distance 
limit of Eq.~(\ref{nn-infty}) we show $V_{CP,nn}$ in Fig.~\ref{fig:Vcp_nn02} 
multiplied by 
different powers of $r$. 
We use parameters from Set 3, which illustrates 
well the qualitative behavior of the other sets. 
The red dotted curve is the CP potential multiplied by $s\,r^6$, where 
$s=100\,\mathrm{fm}$ to fit in the figure. 
The blue dashed and black solid lines stand for 
the dynamic and static polarizabilities versions of $V_{CP,nn}$ (the latter 
indicated by $V^{*}_{CP,nn}$ in the figure), multiplied 
by $r^7$. 
The red thin solid line is the arctan parametrization~\cite{OCaSuc69}, which
is utilized in 
atomic physics (see, for example, Ref.~\cite{FriJacMei02}), that makes the transition from the $1/r^6$ short-distance 
van der Waals to the asymptotic $1/r^7$ Casimir-Polder behavior~\cite{Arn73}. 

The red dotted curve shows a clear $1/r^6$ behavior at small distances 
up to $\approx 20$~fm, meaning that in this region the integrand of 
Eq.~(\ref{eq:integ_ij}) is nearly constant. This $1/r^6$ plateau may be 
just accidental, since this region is dominated by energies larger than 
used to set our parametrizations (\ref{eq:edip-polariz1}), 
(\ref{eq:mdip-polariz1}). This assertion can be checked via the dominance 
of the exponential factor of Eq.~(\ref{eq:integ_ij}): $r\lesssim 20$~fm 
receives contributions from neutron excitations larger than 
$(2\alpha_0\times 20\,{\rm fm})^{-1}\sim 670$~MeV. 
The physics of the Delta resonance appears at about 
$(2\alpha_0\omega_{\Delta})^{-1}\sim 50$~fm, but is minor since it enters 
mostly via $\beta_n(\omega)$, which is numerically of $\sim 10$\%. 
This way, our results can be considered valid for distances beyond 50~fm. 
On the same reasoning, pion production threshold influences the region 
around 100~fm. From the blue dashed curve, one also notices that the 
large distance behavior (\ref{nn-infty}) is only reached beyond $10^3$~fm, 
dominated by dynamic polarizabilities in the region 
$\omega_{\gamma}\lesssim 10$~MeV. 

The above discussion was concentrated on the electromagnetic polarizabilities of the nucleons. The resulting CP interaction is a consequence of two-photon exchange. It is known, though, that the strong interaction also gives rise to long range vdW interaction, the color vdW, arising from multi-gluon exchange. Such force was considered in the scattering of identical heavy nuclei, such as $^{208}$Pb, \cite{HusLimPat90}, and looked for experimentally \cite{VilMitLep93}. Here we mention that this interaction is similar in structure to the electromagnetic one, and can be lumped together.\\

Considerations of the consequences of the interaction potential power laws 
$r^{-4}$, $r^{-5}$, $r^{-6}$ and $r^{-7}$  involving neutrons scattering 
from heavy nuclei were given in Ref. \cite{Pok00}, guided by the work of 
Thaler \cite{Tha59}. 
In the following section we give an account of the influence of our calculated CP interaction in the nn and np systems on the low-energy n-nucleus scattering as done in \cite{Pok00,Tha59}. We do this for the purpose of completeness and to obtain insights into the way the neutron interacts with the constituents of walls of material which are potentially used in neutron confinements in bottles. The full n-wall and wall-n-wall interactions and potentials will be discussed in the  following sections.

\section{Comparison of the CP effect in the nn and pn scatering systems}
The effect of the long range interaction on the neutron-neutron (nn) 
scattering can be estimated using first order perturbation theory.  We can write,
\begin{equation}
f_{nn} (q)= - a_{nn} + f_{nn}^{\infty}(q)
\end{equation}
where $a_{nn}$ is the nn scattering length $a_{nn} = - 18.9 \pm 0.4$~fm.
Taking for the neutron wave function a plane wave, 
$\phi(\bf{k}; \bf{r})$ = $\frac{1}{(2\pi)^{3/2}}\exp{(i\bf{k}\cdot\bf{r})}$, 
the change in the scattering amplitude arising from Eq.~(\ref{nn-infty-num}) is then 
\begin{align}
f_{nn}^{\infty}(q) &= -\left(-0.49 \times 10^{-3}\,{\rm fm^7\cdot MeV}\right)\frac{2\pi^{2}\mu_{nn}}{\hbar^2}\int d\textbf{r} \phi^{\star}(\textbf{k}^{\prime}; \textbf{r})\frac{1}{r^7}\phi(\textbf{k}; \textbf{r})\nonumber\\
& = \left(0.49 \times 10^{-3}\,{\rm fm^7\cdot MeV}\right)\frac{M_{n}}{8\pi\hbar^{2}}\int d\textbf{r}e^{i\bf{q}\cdot \bf{r}}\frac{1}{r^7}
\end{align}
where $q$ = $|\textbf{k}^{\prime} - \textbf{k}|$ = $2k\sin{(\theta/2)}$ is the momentum transfer divided by $\hbar$,  $\mu_{nn}=\textstyle{\frac{1}{2}}M_n$ is the reduced mass, and $M_n$ is the neutron mass. The integral over $\textbf{r}$ can be performed easily \cite{Pok00}. The lower limit of the $r$ integral is set at $r = R$, where $R$ is a radius that characterizes the strong nn interaction and $V_{nn}^\infty (r \le R) = 0$. We have,
\begin{equation}
f_{nn}^{\infty}(q) = \left(0.49 \times 10^{-3}\,{\rm fm^7\cdot MeV}\right)
\left(\frac{M_{n}}{2\hbar^{2}}\right)\frac{1}{R^{4}}F_{7}(q)
\end{equation}
where
\begin{align}
F_{7}(q) & = \frac{\sin{(qR)}}{5qR}+\frac{\cos{(qR)}}{20} -\frac{qR\sin{(qR)}}{60} - \frac{(qR)^{2}\cos{(qR)}}{120}\nonumber\\
&+ \frac{(qR)^{3}\sin{(qR)}}{120} + \frac{(qR)^4}{120}\int_{qR}^{\infty} dt \frac{\cos{t}}{t}
\end{align}
which gives to leading order in $qR$ the following,
\begin{align}
F_{7}(q) &= \frac{1}{4}- \frac{1}{12}(qR)^2\nonumber\\
& + \left[\frac{137}{7200} - \frac{\gamma}{120} -\frac{1}{120} \ln{(qR)}
\right](qR)^4  + O((qR)^6),
\end{align}
where $\gamma\approx 0.5772$ is Euler-Mascheroni constant.

The cross section is given by $|-a + f_{nn}^\infty|^2$. Neglecting the term $|f_{nn}^\infty|^2$, we obtain,
\begin{equation}
\sigma_{nn}(q) = a^2 -2a f_{nn}^{\infty}(q)
\end{equation}

A similar analysis can be performed for the proton-neutron (pn) system, using Eq.~(\ref{np-infty-num}). 
The amplitude is then given by
\begin{equation}
f_{pn}(q) = -a_{pn} + f_{pn}^{\infty}(q).
\end{equation}
Here, $a_{pn}$ is the pn scattering length given by $a_{pn} = - 23.74$~fm. 
The correction owing to the long range interactions is to leading order in $qR$ given by
\begin{align}
f_{pn}^{\infty}(q)& =  \left(0.91 \times 10^{-3}\,{\rm fm^4\cdot MeV}\right) 
\frac{\mu_{pn}}{4\pi\hbar^2}
\int d\textbf{r} e^{i\textbf{q}\cdot \textbf{r}} \left[\frac{1}{r^4} 
- \frac{0.40\,{\rm fm}}{r^5}\right]\nonumber\\
& \approx \left(0.91 \times 10^{-3}\,{\rm fm^4\cdot MeV}\right) 
\left(\frac{M_n}{2\hbar^2}\right)\frac{1}{R}\left[ F_{4}(q) 
- \frac{0.40\,{\rm fm}}{R} F_{5}(q)\right],
\end{align}
where $\mu_{pn}$ is the reduced mass of the proton and neutron and 
where the functions $F_{4}(q)$ and $F_{5}(q)$ are given by \cite{Pok00,Tha59}
\begin{equation}
F_{4}(q) = 1 - \frac{1}{4}\pi qR +\frac{1}{6} (qR)^2 + ....
\end{equation}
and \cite{Pok00}
\begin{equation}
F_{5}(q) = \frac{1}{2} - \left[\frac{11}{36} - \frac{\gamma}{6} 
- \frac{\ln(qR)}{6}\right](qR)^2 +....
\end{equation}

The above results can be summarized by introducing effective scattering lengths for the nn and the np systems. Using the definition 
$a_{\mathrm{eff.}} = a - f(0)$ we find,
for the effective CP-modified scattering length for the $nn$ system,
\begin{equation}
a_{nn, \mathrm{eff.}} = a_{nn} - f_{nn}^{\infty}(0) = a_{nn} 
- 1.23 \times 10^{-4}\left(\frac{M_{n}}{2\hbar^{2}}\right)\frac{1}{R^{4}},
\end{equation}
and similarly for the $np$ system,
\begin{equation}
a_{pn, \mathrm{eff.}} = a_{np} - f_{np}^{\infty} (0)= a_{np} 
+ 0.91 \times10^{-3}\left(\frac{M_n}{2\hbar^2}\right)\frac{1}{R^{2}}
\left(R-0.20\,{\rm fm}\right),
\end{equation}
It is clear that the effect of the CP interaction is more pronounced in the np system; basically an order of magnitude stronger. This becomes clear when calculating the relative effect on the corresponding cross sections.
This discussion about the effect of the Casimir-Polder interaction on the scattering lengths of  nucleon-nucleon system could be of use in the study of charge symmetry violation in hadron physics~\cite{Miller:2006tv}.

It is now a simple undertaking to compare the nn and the pn long-range corrections to the cross sections,
 \begin{align}
\frac{ \Delta\sigma_{nn}(q)}{\Delta\sigma_{pn}(q)} 
&= \frac{a_{nn}f_{nn}^{\infty}(q)}{a_{pn}f_{pn}^{\infty}(q)} \nonumber\\
&= \frac{0.49\,{\rm fm^7\cdot MeV}}{0.91\,{\rm fm^4\cdot MeV}}
\frac{18.9\,{\rm fm}}{23.74\,{\rm fm}}\left[\frac{F_{7}(q)/R^3}
{F_{4}(q) - F_{5}(q)\times0.40\,{\rm fm}/R}\right].
\end{align}
Then, 
\begin{equation}
\frac{ \Delta\sigma_{nn}(q)}{\Delta\sigma_{pn}(q)} 
 \approx \frac{0.43\,{\rm fm^3}}{R^3} \left[
\frac{ \frac{1}{4}-\frac{1}{12}(qR)^2 }
{ 1 - \frac{1}{4}\pi qR + \frac{1}{6}(qR)^2-\frac{0.4\,{\rm fm}}{R}
\left[\frac{1}{2}-\left(\frac{11}{36}-\frac{\gamma}{6}-\frac{\ln(qR)}{6}\right)
\right](qR)^2}
\right].
\end{equation}
Taking for $qR$ the value $1\times 10^{-3}$, with $R = 20$~fm and 
$q=k=5\times 10^{-5} (\mathrm{fm})^{-1}$, corresponding to center of mass 
nn energy of 1~eV, we can obtain the following numerical estimate.
\begin{equation}
\frac{ \Delta\sigma_{nn}(q)}{\Delta\sigma_{pn}(q)} \approx 1.36\times 10^{-5} .
\end{equation}

The estimate given above clearly indicates that at very low energies, the np system is much more influenced by the CP interaction than the nn system. Individually, however, both are very little affected by this interaction when discussing neutrons in containers, such as bottles, at energies in the neV region (ultra cold neutrons). The neutrons feel an over all repulsive interaction with the walls of the container arising from the Fermi pseudo potential which becomes operative when a critical neutron energy is reached \cite{FerMar47, Zel59}. This critical energy varies in value with the material of the wall, but in general it is in the 100's of neV (e.g. for nickel the critical Fermi energy is 252 neV). In containers with walls of aluminum the Fermi pseudo energy or potential is much lower, about 54  neV, corresponding to neutron velocity of 3-24 m/s. 
Therefore the CP effect which is repulsive for the pn system, the dominant constituent in the neutron-wall interaction, has an extremely small effect when compared to the dominant Fermi repulsion.

\section{The neutron-wall interaction}
In discussing the confinement of neutrons inside containers or bottles, one is bound to consider the interaction of neutrons with the wall of the container. The case of an atom and a perfectly conducting wall was considered by Casimir and Polder, and they obtained the following expression valid for very large $r$
\begin{equation}
\label{eq:Vcp-nw1}
V_{CP,aW}(r) = -\frac{3}{8\pi}\hbar c \alpha_{d}(0)\frac{1}{r^4}\,,
\end{equation}
where $\alpha_{d}(0)$ is the dynamic polarizability of the atom at zero frequency. The above formula has been re-derived by many authors and a more general expression was obtained which gives the above as the limiting case as $r \rightarrow \infty$, and a $\frac{1}{r^3}$ form for smaller values of $r$. 
For neutron-wall interaction a similar expression holds and it can be 
written as \cite{khar97}
\begin{equation}
\label{eq:Vcp-nw2}
V_{CP,nW}(r) = -\frac{\alpha_0}{4\pi r^3}J_{nW}(r)
\end{equation}
where 
\begin{eqnarray}
\label{eq:integ_nW}
&&J_{nW}(r)=\int_0^{\infty}d\omega\,e^{-2\alpha_{0}\omega r}
\alpha_{n}(i\omega)Q(\alpha_{0}\omega r)\,,
\nonumber\\[3mm]
&&Q(x)=2x^2+2x+1\,.
\end{eqnarray}
We deduce the neutron-wall interaction based on analogy with the atom-wall 
interaction describing the long-range potential between a neutral polarizable 
particle and a wall. 

Similar to the neutron-neutron case, in the static limit the integration 
above can be done analytically, leading to 
\begin{equation}
\label{nw-infty-num}
V_{CP,nW}^{*}(r)=-\frac{3\alpha(0)}{8\pi r^4}\,,
\end{equation}
which is the asymptotic limit for large distances~\cite{khar97}, 
similar to Eq.~(\ref{eq:Vcp-nw1}). 

\begin{figure}[tbh]
\begin{center}
\includegraphics[width=0.65\textwidth,clip]{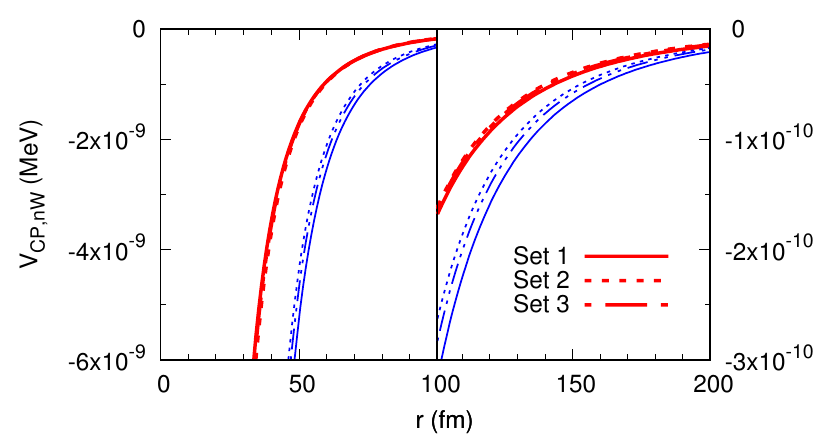}
\caption{\protect 
CP-interaction for a neutron and a wall, as a function of the separation 
distance $r$. Notation is the same as Fig.~\ref{fig:Vcp_nn01}.}
\label{fig:Vcp_nw01}
\end{center}
\end{figure}
\begin{figure}[tbh]
\begin{center}
\includegraphics[width=0.55\textwidth,clip]{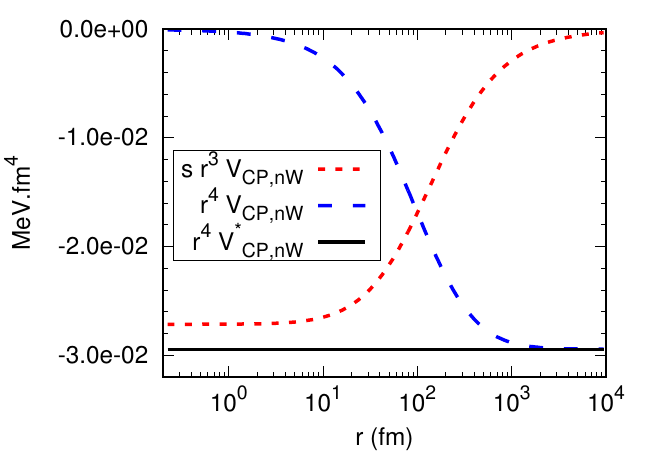}
\caption{\protect 
CP-interaction for a neutron and a wall, as a function of the separation 
distance $r$, multiplied by $s\,r^3$ (red dotted line, with $s=100$~fm) 
and $r^4$ (blue long-dashed line). 
The black solid line is the CP-potential from 
the static limit of the dipole polarizabilities, multiplied by $r^4$.}
\label{fig:Vcp_nw02}
\end{center}
\end{figure}

Figs.~\ref{fig:Vcp_nw01} and~\ref{fig:Vcp_nw02} show the CP-interaction for 
a neutron and a wall, as a function of the separation distance $r$. 
All the qualitative discussion presented for the CP-interaction between two 
neutrons also applies here. Notice that in Fig.~\ref{fig:Vcp_nw02} the 
factors multiplying $V_{CP,nW}$ are $r^3$ and $r^4$. 
The only additional comment is that $V_{CP,nW}$ reaches the expected 
asymptotic behavior slightly faster than $V_{CP,nn}$, most likely due to 
the smaller degree of the polynomial $Q(x)$ compared to $P_{E,M}(x)$
multiplying the polarizabilities, see, respectively, 
Eqs.~(\ref{eq:integ_nW}) and (\ref{eq:integ_ij}).

For very low energy neutrons, in the ultracold region $E_n \approx$ several hundreds of neV, the attractive CP interaction would compete with the repulsive Fermi pseudo potential which is given by $V_F = \frac{2\pi \hbar^2}{M}\rho a$, where $\rho$ is the number density of the atoms in the wall and $a$ is the scattering length of the neutron-nucleus system. The value of $V_F$ depends on the material of the wall. E.g. for Ni, $V_F$ = 252 neV. Accordingly, for neutron energies below this value, there is an overall repulsion from the wall. In the presence of the CP attractive interaction this situation could potentially change.

The final result, which we consider relevant for this work, is the case of a 
neutron between two walls. The result is known for neutral atoms and we 
merely extend it to neutrons. Consider two walls separated by a distance 
$L$ and a neutron at a distance $z$ from the midpoint within the confines of 
the two walls $(-L/2\leq z\leq +L/2)$. The confined neutron is subjected to 
a potential whose form for any value of $L$ is known \cite{khar97}, 
\begin{align}
V_{CP,WnW}(z, L) &= -\frac{1}{\pi L^3}\int_{0}^{\infty}dt\frac{t^2 \cosh{(2tz/L)}}{\sinh{(t)}}\int_{0}^{\frac{t}{\alpha_{0}L}}d\omega \alpha(i\omega) 
\nonumber\\
& + \frac{\alpha_{0}^2}{\pi L}\int_{0}^{\infty}d\omega \omega^{2} \alpha(i\omega)\int_{\alpha_{0}L\omega}^{\infty}dt\frac{e^{-t}}{\sinh{(t)}}
\nonumber\\
&= -\frac{1}{\alpha_{0}\pi L^4}\int_{0}^{\infty}u^3du\,
\alpha\left(i\frac{u}{\alpha_{0}L}\right)
\int_{1}^{\infty}\frac{dv}{\sinh(uv)}\left[
v^2\cosh\left(\frac{2z}{L}uv\right)-e^{-uv}\right]
\label{eq:Vwnw01}
\end{align}
where the latter form is most suitable for numerical calculations, as well as 
deriving analytic results for specific limits. In particular, if one takes 
the static limit of Eq.~(\ref{eq:edip-polariz1}) 
the integrals above can be done exactly, leading to 
\begin{align}
V_{CP,WnW}^{*}(z,L) &= -\frac{\alpha_{n}(0)}{\alpha_{0}\pi L^4}\left\{
\frac{3}{8}\left[\zeta\left(4,\frac{1-f}{2}\right)
+\zeta\left(4,\frac{1-f}{2}\right)\right]-\frac{\zeta(4,1)}{4}\right\}
\nonumber\\
&= -\frac{\pi^{3}\alpha_{n}(0)}{\alpha_{0}L^4}\left[
\frac{3 - 2\cos^{2}(\pi f/ 2)}{8\cos^{4}(\pi f/ 2)}
- \frac{1}{360}\right],
\label{eq:Vwnw02}
\end{align}
where $f=2z/L$ and 
\begin{equation}
\zeta(a,b)=\sum_{k=0}^{\infty}\frac{1}{(k+b)^a}
\end{equation}
is the generalized Zeta function. Eq.~(\ref{eq:Vwnw02}) is nothing but the 
$L\to\infty$ limit~\cite{khar97}, explicitly showing its $L^{-4}$ behavior. 
At the midpoint ($z\to0$) one has 
$V_{CP,WnW}(0,L)=-11\pi^3\alpha_{n}(0)/(90\alpha_{0}L^4)$. If the neutron is 
close to one of the walls, the potential diverges towards negative values. 
\begin{figure}[tbh]
\begin{center}
\includegraphics[width=0.45\textwidth,clip]{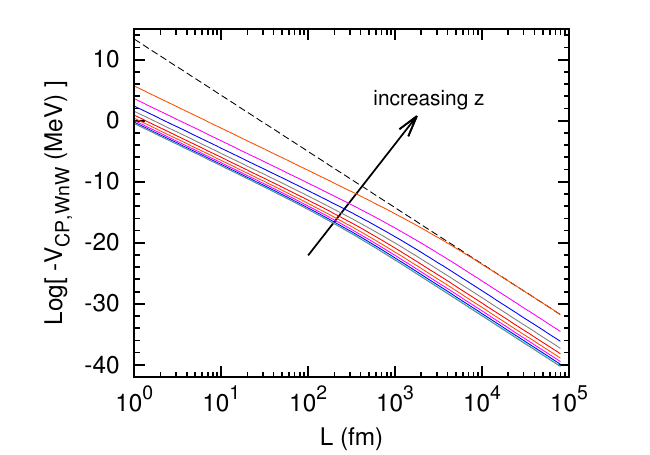}
\includegraphics[width=0.45\textwidth,clip]{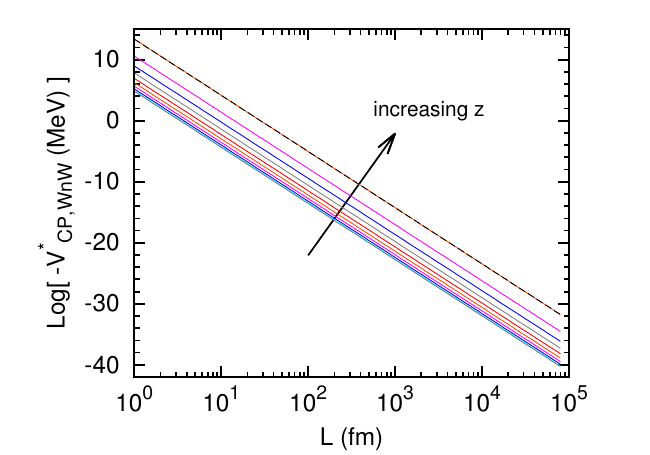}
\caption{\protect 
CP-interaction for a neutron between two walls, as a function of the 
separation $L$. See text for details.}
\label{fig:Vcp_wnw01}
\end{center}
\end{figure}

\begin{figure}[tbh]
\begin{center}
\includegraphics[width=0.45\textwidth,clip]{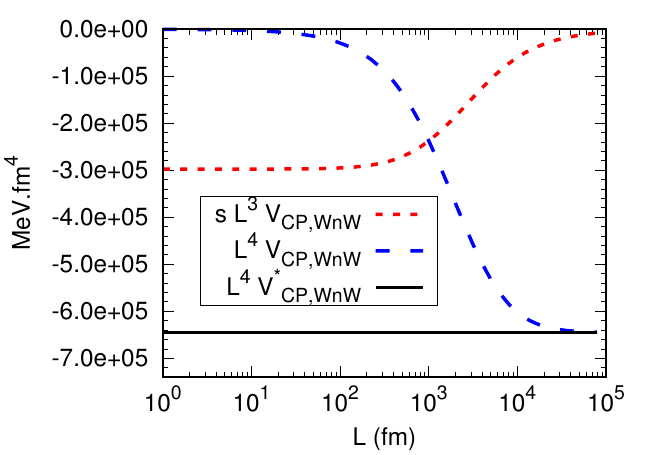}
\includegraphics[width=0.45\textwidth,clip]{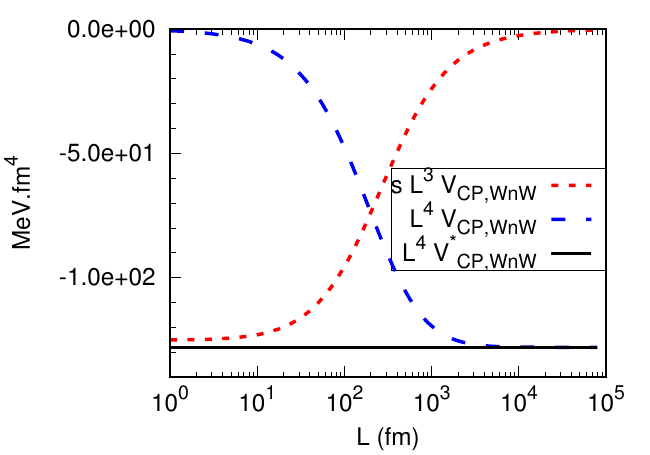}
\caption{\protect 
CP-interaction for a neutron between two walls, as a function of the separation 
$L$ between the walls, multiplied by $s\,L^3$ (red dotted line) and 
$L^4$ (blue long-dashed line). 
The black solid line is the CP-potential from 
the static limit of the dipole polarizabilities, multiplied by $L^4$.
Left panel, $z=0.45L$ and $s=1000$~fm. Right panel, $z=0$~fm and 
$s=200$~fm.}
\label{fig:Vcp_wnw02}
\end{center}
\end{figure}

Fig.~\ref{fig:Vcp_wnw01} shows the numerical results of Eq.~(\ref{eq:Vwnw01}), 
as functions of the separation $L$ between the walls, for several values of 
the neutron distance from the midpoint $z$. 
The lines are for several values of the fraction $f=2z/L$, from 0 to 0.9 in 
steps of 0.1. The left panel, $V_{CP,WnW}$, shows contributions from the 
dynamic electric polarizability of the neutron, while the right panel, 
$V^{*}_{CP,WnW}$, only the static limit. 
The black dashed line on both panels is the result of Eq.~(\ref{eq:Vwnw02}) 
for $f=0.9$ and is drawn just to guide the eye. 
We can check that the static limit is reached only at distances as large as 
$\sim 10^4$~fm, just a tenth of typical atomic dimensions. 
This can be better visualized in Fig.~\ref{fig:Vcp_wnw02}, with analogously 
Figs.~\ref{fig:Vcp_nn02} and \ref{fig:Vcp_nw02}.
Similar to neutron-wall, at small ($\lesssim 10$~fm) 
and moderate ($\sim 100$~fm) distances the behavior resembles more a 
$1/L^3$ falloff than the asymptotic $1/L^4$. The region of this behavior 
is slightly $z$-dependent, as one compares the the left panel ($f=0.9$) 
with the right panel ($f=0$) of Fig.~\ref{fig:Vcp_wnw02}.

\begin{figure}[tbh]
\begin{center}
\includegraphics[width=0.45\textwidth,clip]{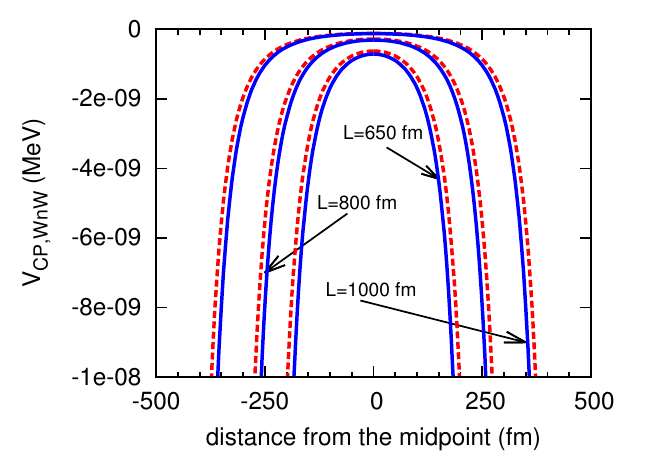}
\includegraphics[width=0.45\textwidth,clip]{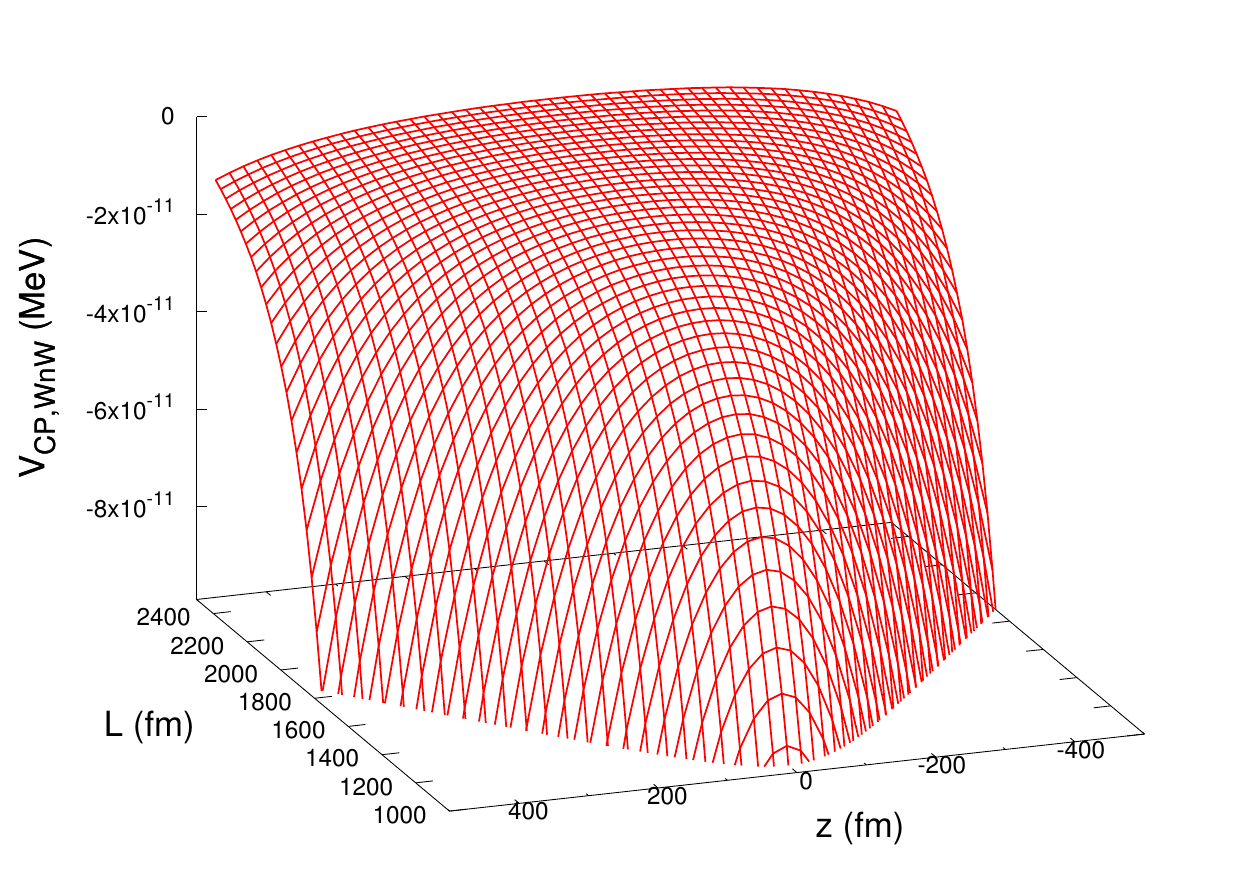}
\caption{\protect 
CP-interaction for a neutron between two walls. Left panel, as a function of 
the neutron position $z$ for three selected values of $L$. 
Right panel, as a function of both the neutron position $z$ and the separation 
between the two walls $L$.}
\label{fig:Vcp_wnw04}
\end{center}
\end{figure}
In Fig.~\ref{fig:Vcp_wnw04} we present the behavior of $V_{CP,WnW}$ as 
a function of the neutron distance from the midpoint $z$. On the left panel 
we select three values of the distance between the walls, $L$, indicated 
in the figure. The red dashed curves stand for the dynamic polarizability, 
and the blue solid curves, for the static limit. One sees that the strength 
of the interaction, as well as the discrepancy of the dynamic and static 
results, increase as one moves the neutron close to one of the walls. 
Finally, on the right panel one can inspect the dependence of $V_{CP,WnW}$ 
on both variables $z$ and $L$, in the region where both the dynamic and 
static cases are not far from each other.

\section{Conclusions}
In this paper we discussed, derived, and analyzed the dispersive Van der Waals 
and the retarded dispersive Casimir-Polder interactions between two neutrons 
and in the proton-neutron system. We found the effect, though very small 
compared to the by far dominant short range strong interaction, is of 
significance at large distances, and is stronger in the pn than in the nn 
system. 
We further assessed the importance of the low-energy nucleon dynamics, namely, 
the pion-production threshold and the first excited state of the nucleon, 
the $\Delta$-resonance (proton (uud), $ \Delta^{+}$, $J^{\pi} = {3/2}^{+}$, 
I = 3/2, $I_{z}$ = -1/2; neutron (udd), $\Delta^{0}$, $J^{\pi} = {3/2}^{+}$, 
I = 3/2, $I_{z}$ = +1/2). We found that they dominate the region 
$\sim 50\,{\rm fm}\lesssim r\lesssim 10^3\,{\rm fm}$ in the nn system, 
the neutron-wall system and in the wall-neutron-wall system. 
This demonstrates that for $r\gtrsim 50\,{\rm fm}$ 
the only aspect of the internal quark structure of the nucleon is the 
induced electric and magnetic dipole moments of the nucleon, a pure dipole 
stretching of the two down quarks against the up quark in the neutron and 
the two up quarks against the one down quark in the proton. 
However, for distances $r\lesssim 50$~fm the studied Casimir-Polder 
interactions are very sensitive to the electromagnetic 
response of the short-distance quark-gluon dynamics inside the nucleon. 
Relevance of our work to confining neutrons inside bottles is briefly discussed.

Our study is exploratory and complementary to the work
by Spruch and Kelsey~\cite{SprKel78} 
for long-range potentials arising from two-photon
exchange in atomic systems.
Spruch and Kelsey replaced the static polarizabilities for two atoms
appearing in the long-range Casimir-Polder potential
by their dynamic polarizabilities. This ansatz was verified rigorously
subsequently by  two
independent Coulomb-gauge quantum electrodynamics calculations~\cite{BabSpr87,Au88}
and shown to agree with the dispersion theoretic formalism
result~\cite{FeiSucAu89}.
Spruch's approach is \textit{significantly} easier to apply than the
formal dispersion theoretic analysis and, at least
as far as practical calculations, it yields correct long-range interaction potentials.
Whether or not this ansatz is good enough or strictly valid for the neutron could be arguable.
For example, we note that in their book, Rauch and Werner (Sec. 10.11, p. 313) state that neutrons ``...provide
the advantage that their Casimir or van der Waals forces are small or perhaps non existing"~\cite{RauWer15}.

We  supposed that the neutron has dynamic electric and
magnetic polarizabilities, for which there is certainly evidence
from dispersion relations and chiral effective field theory calculations 
to Compton scattering, and that the neutron interacts as a polarizable particle.
Moreover, in using our calculated potentials to model an experiment,
other interactions may enter (e.g. the response of the neutron
to an applied magnetic field or (as we discussed) perhaps strong interactions).
The present work suggests that the topic is deserving of further study and experimental investigations.

\section{Acknowledgements}
RH appreciates  conversations with Vladimir Pascalutsa. 
JFB is supported in part by the U. S. NSF through a grant for the Institute of
Theoretical Atomic, Molecular, and Optical Physics at Harvard University and Smithsonian
Astrophysical Observatory.
RH and MSH are supported in part by the Funda\c c\~ao de Amparo \`a Pesquisa do Estado de
 S\~ao Paulo (FAPESP). MSH is also supported by the
Conselho Nacional de Desenvolvimento Cient\'ifico e Tecnol\'ogico  (CNPq). and by the Coordena\c c\~ao de Aperfei\c coamento de Pessoal de N\'ivel Superior (CAPES), through the CAPES/ITA-PVS program.

\end{document}